\documentclass[12pt,preprint]{aastex}

\shorttitle{A Shell and a Jet in RS Oph}
\shortauthors{Rupen, Mioduszewski, \&\ Sokoloski}

\begin{document}


\title{An Expanding Shell and Synchrotron Jet in RS Ophiuchi}


\author{Michael P.\ Rupen and Amy J.\ Mioduszewski}
\affil{National Radio Astronomy Observatory, 1003 Lopezville Road,
Socorro, NM 87801}
\email{mrupen@nrao.edu, amiodusz@nrao.edu}

\and

\author{Jennifer L.\ Sokoloski\altaffilmark{1}}
\affil{Columbia Astrophysics Laboratory, 550 W.\ 220th Street, 1027 
Pupin Hall, Columbia University, New York, NY 10027}
\email{jeno@astro.columbia.edu}


\altaffiltext{1}{National Science Foundation (NSF) Astronomy and
Astrophysics Fellow.}


\begin{abstract}
We report 
high-resolution radio imaging of the recurrent nova RS Ophiuchi (RS
Oph) during the first month of the 2006 outburst, using the Very Long
Baseline Array (VLBA).  Observations made on days 20.8 and 26.8 of the
outburst show a synchrotron-emitting partial shell that is much
brighter to the east than to the west.  Assuming the broad component of the
infrared lines corresponds to the outermost part of the shell seen by the
VLBA, the distance to the source is $2.45\pm0.4$\,kpc.
The circular shape and spectral indices of the shell
emission challenge simple models for the radio structure
immediately after the outburst.  The second epoch also shows
an additional, resolved, synchrotron-emitting component well to the
east of the shell.  Its inferred velocity is
comparable to the escape speed from the surface of a high-mass white
dwarf.  This component was not seen in the first epoch.  Its
appearance may be related to the outflow reaching the edge of the
nebula created by the red giant wind, which had been re-filling the
void left by the last outburst in 1985.  This
eastern component is likely related to the jets previously seen in
this and other symbiotic stars, and represents the earliest clear
detection of such a jet, as well as the best case yet for synchrotron
emission from a white dwarf jet.
\end{abstract}


\section{Introduction}

The most recent major outburst of the symbiotic star and recurrent
nova RS Ophiuchi (\object{RS Oph}) was discovered near the optical
peak on 2006 February 12.83 \citep{Nar06}, hereafter taken as time
zero.
RS Oph is a binary system 
in which a red giant deposits material
through its wind onto a white dwarf (WD) companion \citep[see,
e.g.,][]{ken86}.  Approximately every 20 years, the system erupts as a
nova, as 'runaway' thermonuclear burning in the hydrogen-rich
white-dwarf envelope leads to a visual brightness increase of more
than 6 mag in less than 2 days, a sustained period with WD luminosity
near the Eddington limit, and the ejection of material from the WD
surface \citep[e.g.,][]{Bod87}. The last such eruption was in 1985.
The frequency of these outbursts suggests that the mass of the WD in
RS Oph is very close to the Chandrasekhar limit
\citep[e.g.,][]{Hac01,Sok06e}, making this a prime candidate for a
Type Ia supernova progenitor system.

Whereas most novae detected in the radio emit via thermal free-free
bremsstrahlung
from the shell of ejecta \citep[e.g.,][]{Sea89}, post-outburst radio
emission from RS Oph instead emanates primarily from the dense
circumstellar material (CSM) that is swept up and shocked by the
fairly low-mass
ejected shell.  Observations after the previous outburst in 1985 with
the European Very Long Baseline Interferometry Network (EVN), the
Multi-Element Radio Linked Interferometer Network (MERLIN), and the
Very Large Array (VLA) suggested that the radio emission was initially
non-thermal at low frequencies, with a possible thermal component
becoming dominant later in the outburst. The radio source
consisted of a resolved central component plus possible bi-polar jets
expanding at 1.5 mas/day \citep{pad85,Hje86,Tay89,dav87,por87}.

Although jets have been observed from an increasing number of
accreting white dwarfs in symbiotic stars and supersoft X-ray sources
\citep{Sok04}, and even several novae \citep{Kaw06,Iij03}, these jets
are generally not spatially well resolved, and basic questions such as
the relationship between jet production and outbursts remain.  Most
jets from white dwarfs have velocities of hundreds to thousands of km
s$^{-1}$, and when jet emission can be spatially isolated, it usually
appears to be thermal \citep[e.g.,][]{Hol97,Kel01,Bro04,Gal04}.  In a
few cases, however, observations have hinted that particles can be
accelerated sufficiently that non-thermal radio and/or X-ray emission
can also be produced \citep{Cro01,Nic07}.

Here we report the detection of an expanding, synchrotron-emitting
ring just weeks after the nova explosion, and, between three and four
weeks after the outburst, the appearance of a synchrotron-emitting
feature to the east of this ring aligned with the putative jet from
1985.
We describe our observations of the expanding ring and the eastern component in 
\S2.  We discuss these observations in more detail in \S3, and summarize our
conclusions in \S4.

\section{Observations}
We observed RS Oph with the National Radio Astronomy
Observatory's\footnote{The National Radio Astronomy Observatory is a
facility of the National Science Foundation operated under cooperative
agreement by Associated Universities, Inc.} (NRAO's) Very Long
Baseline Array (VLBA) under project code BS167 in two four-hour runs,
on 2006 March 5 and 2006
March 11, corresponding to days 20.8 and 26.8 of the outburst,
respectively.  The first run used all antennas but Hancock, while the
second used the entire 10-antenna array.  We observed the source in
two frequency bands, consisting of 64\,MHz in each of the two dual
polarizations, centered on 1.6675 and 4.9875\,GHz, recorded at a rate
of 256 megabits per second.  To remove atmospheric and instrumental
effects, we employed phase referencing using the nearby calibrators
J1745$-$0753 (1\fdg67 away; primary calibrator at 5.0\,GHz) and
J1743$-$0305 (3\fdg27 away; primary calibrator at 1.7\,GHz).  The position
of the former was taken from the Second VLBA Calibrator Survey
\citep{Fom03}, with a quoted (rms) accuracy of 0.50\,milliarcseconds;
the position of the latter was taken from the International Celestial
Reference Frame \citep[ICRF;][]{Fey04}, with a quoted accuracy of
0.77\,milliarcseconds.
We
correlated all four polarization pairs using the VLBA correlator in
Socorro, NM, and calibrated, imaged, and self-calibrated the data in
the standard fashion using NRAO's Astronomical Image Processing System
\citep[AIPS;][]{Gre03}.  Phase referencing aligned the 5.0\,GHz images
to better than 0.5\,mas.  The 1.7\,GHz images differed by
$\sim8$\,mas, rather larger than expected from residual ionospheric
effects, but not alarmingly so, given the poor weather during the
first observations.  We shifted the lower frequency images to align
the centers of the rings with those seen at 5.0\,GHz.

Figure~\ref{fig:images} shows the total intensity images.  To a limit
of a few percent (or the rms noise, whichever was higher), no linear
or circular polarization was detected in either epoch, at either
frequency.  There may be some missing radio emission on scales much
larger than sampled by the VLBA -- the corrected 5\,GHz flux density
on day 20.8 ($\sim32\rm\,mJy$; see below) is somewhat lower than that
seen by MERLIN at 6\,GHz on the same day \citep[about
$40\rm\,mJy$;][]{OBr06a}.

\subsection{The Expanding Ring\label{obs:expring}}

\begin{figure}
\epsscale{.80}
\plotone{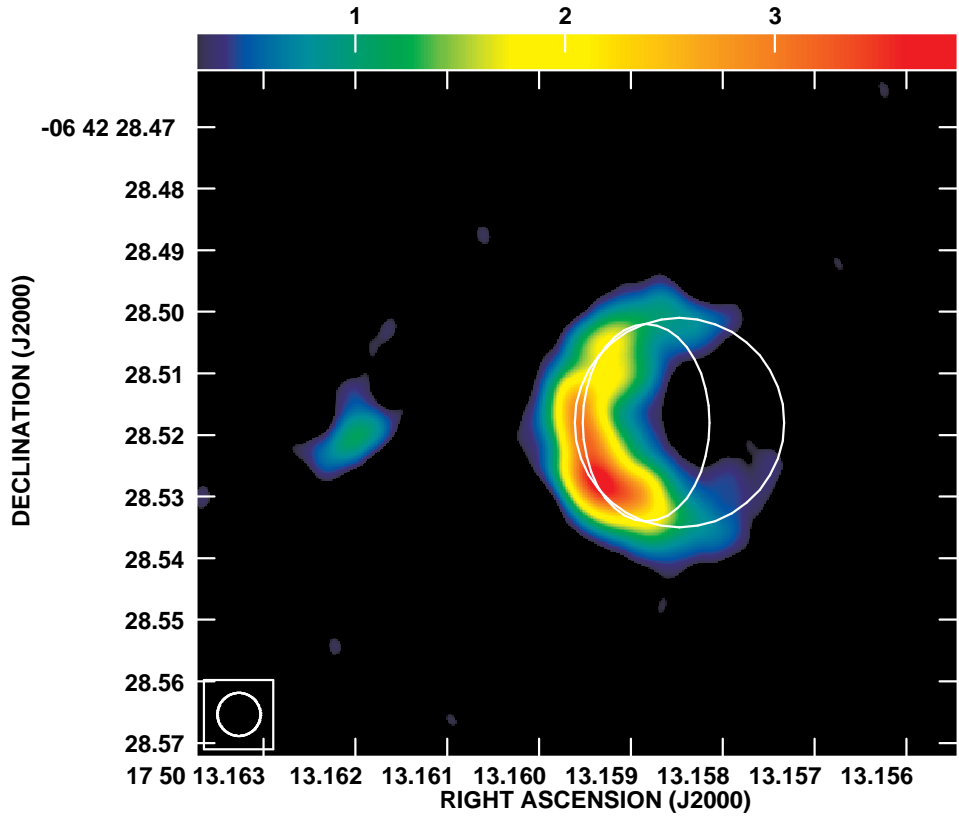}
\caption{Tilted circular rings superposed on the 5.0\,GHz image from day 20.8,
  for inclinations of 0\degr (face-on) and 50\degr, using radii of
  31 and 29\,milliarcseconds, respectively.  The orbital inclination is
  thought to be $\sim50\degr$ \citep{Qui03,Bra07}.  The color scale for the 
  image ranges from  0.25 (black) to 3.865 (red) $\rm mJy\,beam^{-1}$.
  \label{fig:rings}
  }
\end{figure}

All four images are dominated by a ring which is much brighter to
the east than to the west.  Despite this brightness asymmetry, the
ring is consistent with being circular (Fig.\,\ref{fig:rings}).  In
particular, a two-dimensional ring with an inclination corresponding
to that of the binary orbit \citep[$\sim50\degr$:][]{Bra07,Qui03}
provides a significantly worse fit than a face-on ring or sphere. These
results are based on simulated VLBA observations of circular and elliptical
rings of various thicknesses and axis ratios, with brightness distributions
around the ring similar to that observed.  These simulations take account of
such important interferometric uncertainties as the effects of limited
uv-coverage and non-linear processing, including deconvolution in
particular.

\begin{figure}
\plotone{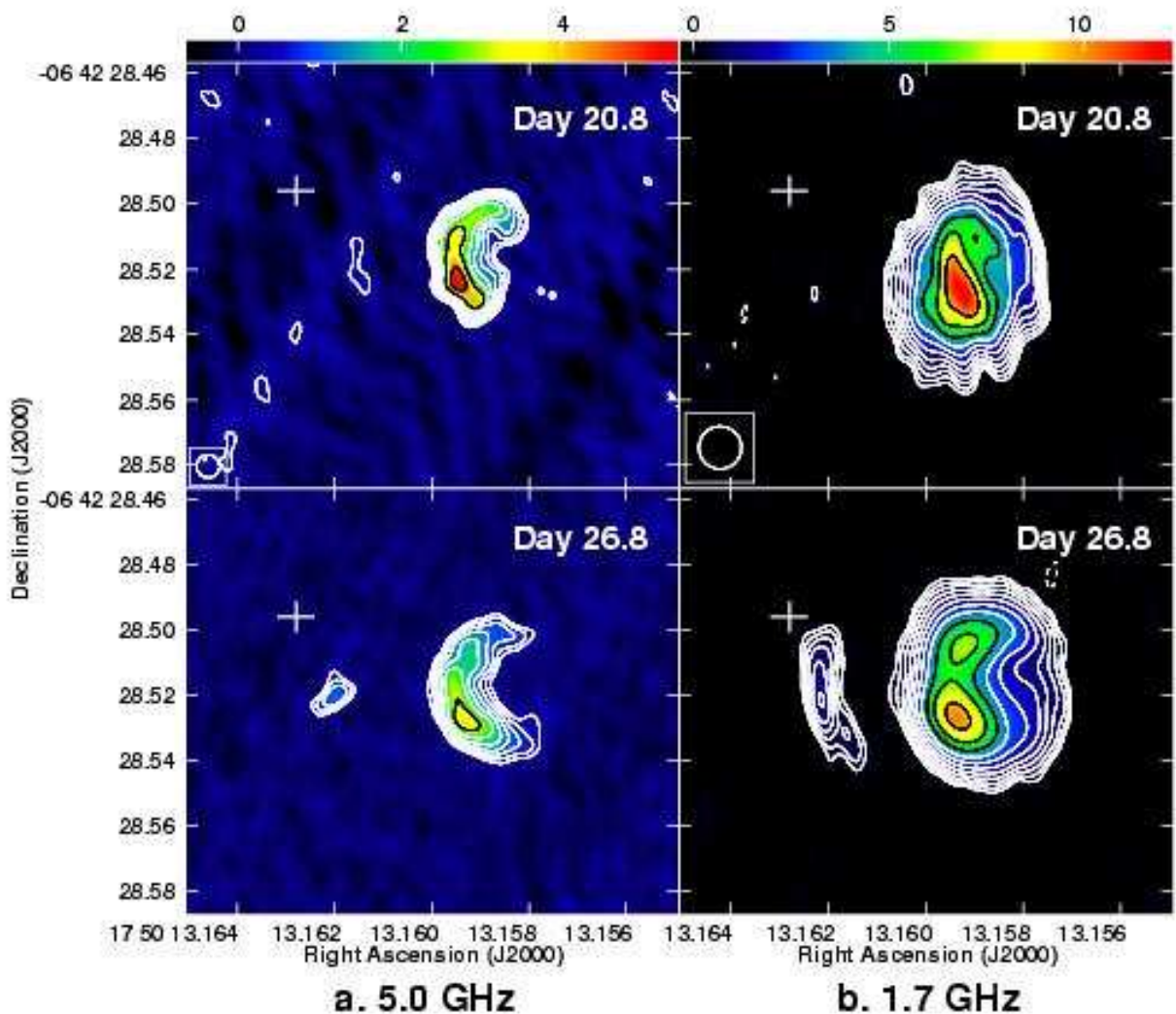}
\caption{VLBA total intensity images of RS Oph, from 2006 March 5 
  (day 20.8; upper) and 2006 March 11 (day 26.8; lower).   
  (a)~5.0\,GHz images (left column), convolved to a resolution of
    $7.6\times7.6\rm\,mas$.  Contours are
    $\pm0.25\,\rm mJy beam^{-1}\times2^{n/2}$, $n=2, 3, \dots$,
  and the color scale ranges from 0 to
    $6\rm\,mJy\,beam^{-1}$.
  The peak flux densities in the two images are
    $5.9\pm0.25\rm\,mJy\,beam^{-1}$ on day 20.8, and
    $4.3\pm0.11\rm\,mJy\,beam^{-1}$ on day 26.8.
  Note that the 5.0\,GHz images are inherently less sensitive to
  large structures, due to the difference in interferometric baseline
  coverage as measured in wavelengths.  The total 5.0\,GHz flux densities,
  uncorrected for this effect, are 41.3 and 26.3\,mJy.
  (b)~1.7\,GHz images (right column), convolved to a resolution of
    $13\times13\rm\,mas$.  Contours are
    $\pm0.12\,\rm mJy beam^{-1}\times2^{n/2}$, $n=2, 3, \dots$,
  and the color scale ranges from 0 to
    $13\rm\,mJy\,beam^{-1}$.
  The peak flux densities in the two images are
    $13.0\pm0.12\rm\,mJy\,beam^{-1}$ on day 20.8, and
    $10.3\pm0.13\rm\,mJy\,beam^{-1}$ on day 26.8, with the error bar
    reflecting the off-source rms noise.  The total flux densities
    are 43.1 and 30.1\,mJy.
  The
  cross to the northeast in each panel marks the optical position of RS\,Oph, 
  from NOMAD
  (see \S\,\ref{obs:expring}); the size of the cross represents the NOMAD
  error bars.
  \label{fig:images}
  }
\end{figure}

The
ring center (as measured at 5\,GHz on both days) at (J2000) is:
$$17^h 50^m 13\fs1583, -06\degr 42^\prime 28\farcs517,$$
with an uncertainty of at most one milliarcsecond (mas).
We checked for systematic errors in our astrometry by phase referencing our
secondary calibrator (J1743$-$0350) to our primary calibrator (J1745$-$0753)
at 5\,GHz.  The resulting position agreed with that given in the ICRF to
0.4\,mas in right ascension and 0.6\,mas in declination -- well within
the $1\sigma$ error bars reported in the ICRF.  Since
these calibrators are $\sim4\degr$ apart, while RS\,Oph is separated from the
primary calibrator by only $1\fdg67$, systematic errors are far below other
uncertainties in the measurement.  Our VLBI position for RS\,Oph is however
$\sim100$\,mas from the optical position
from the Naval
Observatory Merged Astrometric Dataset \citep[NOMAD;][]{Zac04}:
$17^h 50^m 13\fs1628\pm9\rm\,mas$,
$-06\degr 42^\prime 28\farcs496\pm11\rm\,mas$.
This optical position corrects for proper motion to the epoch 2006.19, and
is shown as a cross in Fig.~\ref{fig:images}.
There is no obvious explanation for the radio/optical offset. 

The width of the ring is barely resolved in the first epoch, and
clearly resolved in the second, suggesting a growing thickness with
time.  We fit the observed ring as a circle with an asymmetric
brightness distribution, thickened by convolution with a symmetric
Gaussian, and then convolved with the beam (point spread function).
The full-widths-at-half-maxima (FWHMs) for the smoothing gaussians
were $2.5^{+1.0}_{-2.5}\rm\,mas$ on the first day, and
$4.5\pm0.5\rm\,mas$ on the second.  The corresponding radii were
$16.25\pm0.5\rm\,mas$ and $17.5\pm0.5\rm\,mas$.  These radii refer to the
mid-points of the annuli; the outermost parts of the rings (mid-point
+ FWHM, from the fits) are at $\sim17.5$ and $\sim19.7$\,mas.  The error
bars quoted here are more nearly $3\sigma$ than $1\sigma$ errors, insofar
as this has meaning for non-Gaussian error distributions, and the difference in
widths is thus likely to be real.

  The asymmetry of the ring is much more pronounced at 5\,GHz than at
1.7\,GHz.  This difference is not an instrumental effect: simulating 5\,GHz
observations of the 1.7\,GHz model shows that the 5\,GHz observations would
have shown the western emission clearly, had it been as
(relatively) strong as at 1.7\,GHz.   
  Thus the spectrum of the eastern part of the ring is {\it intrinsically}
relatively flat, while that of the western part falls sharply with
frequency.  Based on the simulations, we multiplied the
observed flux density at 5\,GHz by a factor 1.07 on day 20.8, and 1.10
on day 26.8, as a first-order correction for the differing uv-coverage
between 5 and 1.7\,GHz.\footnote{Figures\,\ref{fig:images} and
  \ref{fig:rings} do {\it not} reflect this correction, as it is very
  likely to vary with position.}
The resulting flux densities at 1.7 and 5\,GHz are given in
Table~\ref{tab:spind}.

\subsection{The Eastern Component}\label{obs:ec}
  
  Our second set of observations revealed an additional component
$\sim52\rm\,mas$ to the east of the center of the ring.  This new
eastern component was present at both 1.7 and 5.0\,GHz.  The 1.7\,GHz
emission at this location is clearly resolved in the north/south
direction, with a Gaussian fit yielding a rough deconvolved size of
$23.7\pm2.2\rm\,mas$ 
($58.0\pm5.4\rm\,AU$ at 2.45\,kpc [see \S\ref{sec:distance} for a derivation of the distance]).
As with the ring, we simulated 5.0\,GHz observations of the 1.7\,GHz
model, and found that, while the eastern component would appear much
smaller at 5.0\,GHz (as observed),
the source spectrum must be intrinsically quite steep (see
Table~\ref{tab:spind}).

This component was not seen in our first set of observations.
The 1.7\,GHz map shows no source to the east of the ring: the
$5\sigma$ flux density upper limit is $0.6\rm\,mJy\,beam^{-1}$,
compared to the peak flux density on March 11 of
$1.3\rm\,mJy\,beam^{-1}$.  The $3\sigma$ 5\,GHz upper limit is
approximately $1\rm\,mJy\,beam^{-1}$.
Further VLBA observations on 2006 Mar 18 through 29 clearly detect the
eastern component; we postpone discussion of those data to a future
paper.

\section{Discussion}

\subsection{The Ring}

\subsubsection{Expansion Rate and Shape}

\begin{figure}
\plotone{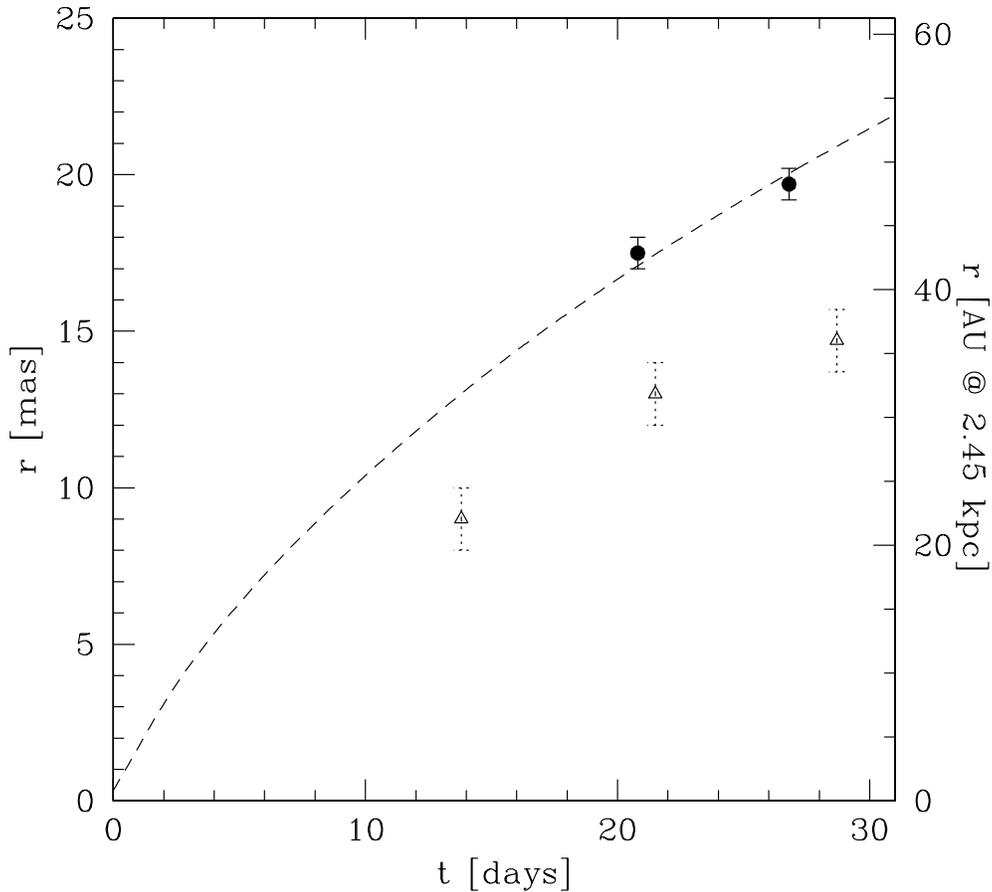}
\caption{The radius of the ring as a function of time since the optical
  peak on 12.83 Feb 2006 UT.  The solid circles represent the outer radii of
  the fit annuli (see text).
  The triangles shows the data of \citet{OBr06c}, assuming a
  1\,mas error bar  (see text); these size estimates are only
  approximate, and the disagreement between the two groups is likely due
  to differences in the methods of their derivations (e.g., it is likely
  that \citet{OBr06c}'s values refer to the mid-point rather than the outer
  edge of the ring).
  The dashed line
  shows the expansion detailed in Table\,\ref{tab:radfits}, scaled
  up by a factor 1.7 to match the width of the broad infrared spectral
  component (see \S\,\ref{sec:distance}).
  \label{fig:exp}
  }
\end{figure}

  The expanding ring is most naturally associated with the shock wave
resulting from the outburst, and seen in the strong early X-ray
emission \citep[e.g.,][]{Sok06d,Osb06a,Bod06b} and broad optical, infrared,
and X-ray lines \citep[e.g.,][]{Bui06,Nes06,Das06b}.
Those data suggest substantial
deceleration, as expected for a shock sweeping up more and more of the
red giant's stellar wind. 
Figure\,\ref{fig:exp} compares the radius thus predicted from the
optical, IR, and X-ray velocities 
with that observed in the radio.  The solid
circles represent outer radius estimates from the current data, derived through
simulations of thick, asymmetric rings as discussed in
\S\ref{obs:expring}.  
The
open triangles represent the radii of \citet{OBr06c}, measured as half
the maximum distance between peaks in a series of north-south slices.
The nominal error bars of \citet{OBr06c} are 0.2\,mas, but, as they
point out, these nominal errors almost certainly underestimate the
true ones, and we plot 1\,mas error bars in the
figure.  The systematic difference between O'Brien et al.'s
measurements and ours, with the former lying some 20\%\ low, is likely
due to the differences in fitting procedures, combined with an
evolving surface brightness distribution and changing angular
resolution \citep[cf.\ \S5.6 of][]{Bar02}.
Nonetheless, the overall trend of radio radius with time, as measured
by either group, matches the X-ray/optical/infrared predictions nicely.  
The agreement between the size of the radio ring and the
expected location of the blastwave also support the conclusion
that, unlike the IR continuum emission, which primarily emanates from a region
within a few mas of the central binary \citep{Mon06,Lan07},
the radio 
emission is associated with
the expanding shock.

  As shown in Fig.\,\ref{fig:rings}, the ring is basically circular.
This geometry is consistent with a spherical shock projected onto the
plane of the sky \citep[see, for instance,][for similar observations
of supernova 1993J]{Bar02} or a two-dimensional shock in the plane of
the sky.  There is no evidence in the ring shape either for an
asymmetric explosion or for any special geometry of the circumstellar
material.  In particular, we see no sign that the ring is inclined
either parallel or perpendicular to the binary orbit
\citep[$\sim50\degr$:][]{Bra07,Qui03}.
  The fractional width of the ring (thickness divided by radius)
increases from $0.15^{+0.06}_{-0.15}$ on day 20.8, to $0.26\pm0.03$ on
day 26.8 ($\sim3\sigma$ error bars).
This breadth suggests that the emission arises in the forward rather than
the reverse shock, since the latter is expected to be much thinner.
The {\it change} in relative width shows that the
evolution is {\it not} scale-free, thus ruling out a common
theoretical simplification \citep[e.g.,][]{Che82}.

\subsubsection{Distance} \label{sec:distance}

  Comparing the angular size observed by VLBI with the radius inferred from
optical, X-ray, and infrared observations gives a direct estimate of the 
distance to the remnant.   Three estimates of the evolving shock veocity
are currently available.

  First, the earliest optical spectra give initial velocities of 
$3500\pm150\rm\,km\,s^{-1}$ \citep[based on][and early spectra reported on
VSNet]{Bui06}, which we take as constant between the explosion (assumed to
occur on Feb.\,$12.6\pm0.15$, or $\Delta t\equiv t-t_0= -0.23\pm0.15$ days
with respect to $t_0$, the time of optical maximum [Feb.\,12.83])
and $\Delta t=1.47$\,days.

\begin{figure}
\plotone{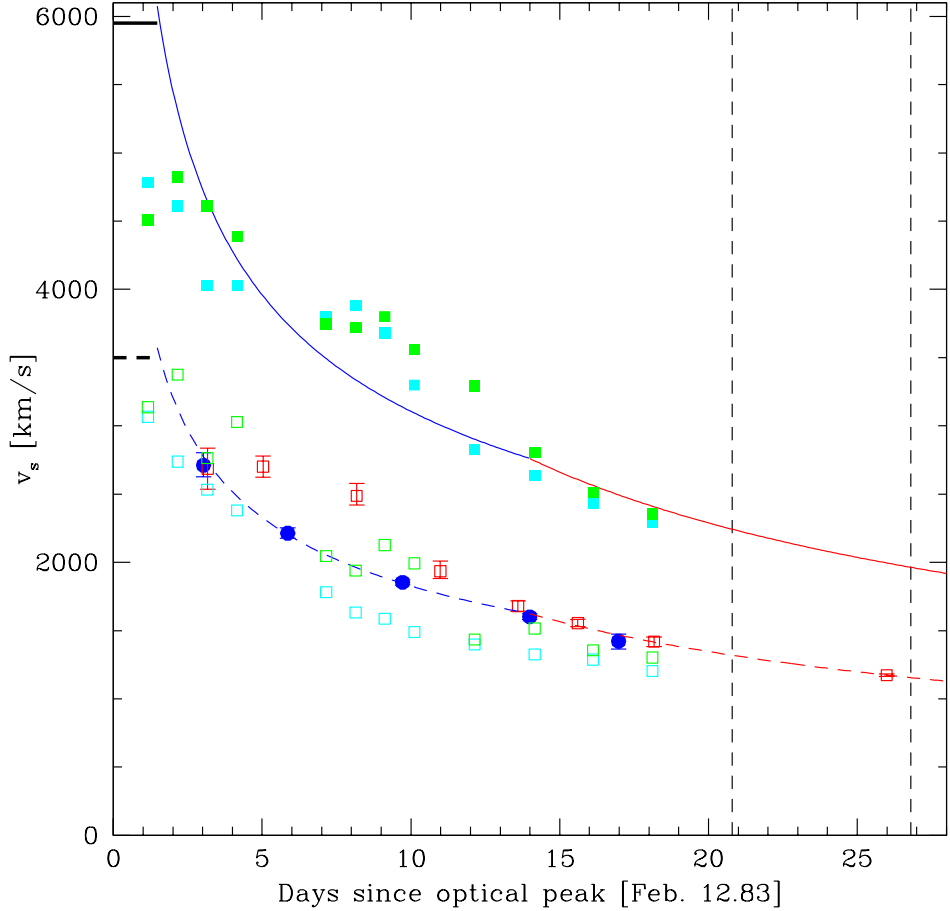}
\caption{The velocity of the shock vs.\ time, as derived from 
  infrared, optical, and X-ray data.   The observed widths of the
  \ion{O}{1}\ line are shown in green, and those of the Pa$\beta$ line
  in light blue (cyan); the open squares represent the full-widths at
  half-maximum, and the solid squares show the half-widths at zero
  intensity (HWZIs).  The blue dots and error bars
  are shock velocities inferred from {\it RXTE} X-ray data in \citet{Sok06e},
  while the red ones are similarly derived from {\it Swift} X-ray data in
  \citet{Bod06b}.  The blue
  and red dashed curves represent power-law fits (see Table~\ref{tab:radfits})
  to the observed X-ray velocities, 
  ignoring data from \citet{Sok06e} taken after day 14.5, and data from
  \citet{Bod06b} taken before day 13.5.   The dashed black horizontal line
  shows the velocity for the earliest part of the
  outburst, taken from early optical reports.  As discussed in
  \S\,\ref{sec:distance},
  the solid black, blue, and red curves show the result of multiplying the
  X-ray and optical fits by 1.7, to match the infrared HWZIs.  These solid
  curves represent our best estimate of the shock velocity.
  The vertical dashed lines show the dates of our VLBA
  observations.
  \label{fig:xraytemp}
  }
\end{figure}

  Second, shock velocities may be derived from the observed
X-ray temperatures, using the standard strong shock approximation and
assuming the standard mean molecular weight ($\mu=0.6$)
for a fully ionized plasma.  The X-ray emission is clearly complex,
with indications of multiple temperatures and densities present at any given
time.  Here we are interested in the
highest temperature gas, corresponding to the material heated by the
decelerating shock shown in the radio images, and adopt single-temperature
fits to the highest energy X-rays seen in a given observation.
The X-ray temperatures at early times, when the
temperatures are highest, are best determined by the {\it Rossi X-ray Timing
Explorer} \citep[RXTE;][]{Sok06e}, which responds most efficiently to
higher-energy photons; at late times, as the shock slows and the
corresponding temperature decreases,  the relatively
soft response of {\it Swift} provides the
more relevant spectral coverage \citep{Bod06b}.  We therefore 
use the
measurements of \citet{Sok06e} 
before day 14.5, and those of 
\citet{Bod06b} 
after day 13.5.
The X-ray velocities are well fit by two independent power laws, as
detailed in Table\,\ref{tab:radfits} and shown by the dashed lines in
Fig.\,\ref{fig:xraytemp}.

  The third and final set of velocity estimates comes from infrared
spectroscopy.  \citet{Das06b} measured the (deconvolved) full-widths at half
maxima (FWHMs) and half-widths at zero intensity (HWZIs) of the observed line
profiles of two infrared lines, \ion{O}{1}\ and Pa$\beta$, from 1.16 to 47.04
days after the explosion.  While the measurements for the two lines are in
rough agreement, and the FWHMs agree with the optical and X-ray velocities
discussed above, the FWHMs for both lines -- and the velocities inferred from
the optical and X-ray data -- are systematically low (by a factor
$\sim1.7$) compared with the FWZI.  This discrepancy was also noted by
\citet{Eva07}.   Fig.\ 2 of that paper shows that the difference between
FWHM and HWZI results from the shape of the line
profile: a strong, narrow, roughly Gaussian peak, surrounded by a faint,
broad plateau.  The narrow component suggests an expanding wind or outflow,
while the broad plateau is more characteristic of a thin shell
\citep[e.g.,][]{Lyn06,Ber87,Gil99}.   The consistent ratio of the FWHMs and
HWZIs, and the roughly constant ratio of the peak flux density in the two
components, shows they remain closely associated for at least the first 40
days after the explosion.
 
  The early optical reports may well have missed a faint, broad line component.
The much lower velocities derived from the X-ray temperatures, and their
agreement with the FWHM of the narrow component of the infrared lines, is more 
puzzling.  \citet{Tat07} suggest that cosmic ray acceleration at the blast 
wave may explain both the velocity discrepancy and the relatively rapid 
deceleration of the shock.  

  Here we assume the broad component of the infrared lines represents the
shock velocity, as both the line shape and the extreme velocity suggest 
a thin, rapidly expanding shell.  The scatter in the infrared measurements 
from day to day suggests some short-lived physical effects, or some
uncertainty in the determinations of the line widths.\footnote{Fig.\ 2 of
  \citet{Das06b} shows error bars which appear sufficient to explain much
  of the scatter, but those error estimates are not given explicitly in the
  text or in Table 2 of that paper.}
We therefore multiply
the smooth fit to the
X-ray and optical velocities (see above) by 1.7, to give a smooth curve
which roughly reproduces the decline in the infrared HWZI.  With these
assumptions, the errors on the shock radii at the times of our VLBA
observations are dominated by the error on the time of the explosion.

  Comparing these radii to the angular sizes corresponding to the outer part
of the shock ($\theta=17.5\pm0.5$\,mas at $\Delta t=$ 20.8\,days;
$\theta=19.7\pm0.5$\,mas 
for $\Delta t=$ 26.8\,days) gives two fairly independent measurements of the
distance:
\begin{eqnarray}
  \nonumber
  D_{20.8}= r_{20.8}/\theta_{20.8}= 2.392\pm0.078\,{\rm kpc}\\
  \nonumber
  D_{26.8}= r_{26.8}/\theta_{26.8}= 2.492\pm0.071\,\rm kpc
\end{eqnarray}
where the error bars are statistical, and dominated by the uncertainties in
$\theta$.
Taking the weighted average of these two distances gives a distance of
$2.45\pm0.05$\,kpc.  
 
  The major uncertainty in this estimate lies in the relation between the
sizes measured with VLBI, and the velocities inferred from other
wavelengths.  First, if the ring is a two-dimensional structure inclined at
an angle $i$ to the line-of-sight, that will increase the quoted distances
by $1/\cos i$: a factor 1.6 for $i=50\degr$.
Given the observed circularity of the ring, 
this is likely to be a $<10\%$ effect. 
Second, we have assumed that the infrared velocities reflect those of the
leading edge of the radio emission.  Adding an additional 10\% (rms) systematic
uncertainty based on the possibility that the outer edge of the observed
radio shock lies within the high-velocity region traced by the extreme
infrared velocities, the final distance estimate is then
$2.45\pm0.05\pm0.37$\,kpc, where the first error
bars refers to the random and the second to the systematic errors.

  Previous distance estimates are summarized by \citet{Bod87}; see also
Barry et al. (in prep.).  These were
based on spectral luminosity determinations, the presence or absence of
absorption features at known distances, and a very rough relation between
the distance and (cold) \ion{H}{1} column density \citep[see
primarily][]{Hje86}.  
\citet{Bod87} concluded that the distance was likely to be
around $1.6\pm0.3$\,kpc.  Our more direct estimate is higher but not
enormously so.  Further, since a significant inclination for the ring can
only {\it increase} our distance estimate, the comparison
implies again that the ring is nearly face-on. 

\subsubsection{Emission and Absorption}

  Given the close agreement between the blast wave velocity evolution
inferred from X-ray, optical, and infrared observations and the
expansion of the radio ring,
it seems worth checking whether a significant fraction of the radio
flux density could be thermal bremsstrahlung from the X-ray-emitting
gas. Fits to the data of \citet{Bod06b} (Table\,\ref{tab:radfits})
give the temperature of the post-shock gas as 
$\sim2.4\times10^7\rm\,K$ on
day 20.8 and $\sim1.8\times10^7\rm\,K$ on day 26.8.  These plasma temperatures
are close to the observed peak surface brightness, which corresponds
to an optically-thick black body of a few times $10^7\rm\,K$.
However, a mass of more than $10^{-5}\,M_\sun$ 
would be required to
make the thermal gas optically thick in the radio, and the
corresponding thermal energy would be more than $10^{44}\rm\,ergs$ --
an order of magnitude above the total energy of the outburst
\citep[e.g.,][]{Hac01,Sok06e} -- with a correspondingly high X-ray
luminosity. Clumping of higher-temperature gas
would reduce the required mass somewhat, but not enough to remove
these difficulties.  Thermal emission is thus unlikely to play a major
role in the few-GHz radio flux at this stage of the outburst.

  The obvious alternative is synchrotron emission, which can easily
explain the observed high surface brightness.  Optically-thin
synchrotron spectra generally range from $\nu^{-0.6}$ to $\nu^{-1}$,
providing a natural explanation for the $\sim\nu^{-0.7}$ spectrum of
the western side of the ring.  The eastern part of the ring, with
$\alpha={-0.20\pm0.02}$ ($S_\nu\propto\nu^\alpha$) on day 20.8 and
$\alpha=-0.27\pm0.01$ on day 26.8, is more puzzling.  Foreground
absorption in a $1/r^2$ medium would be expected to change rapidly
over this time period, while the observed spectrum is fairly constant.
Also, whereas any foreground absorption (or synchrotron
self-absorption) should have caused the flux density to decrease with
decreasing frequency below 1.7\,GHz,
in fact the 0.61\,GHz flux density of $48\pm2$\,mJy on day 20.3
\citep[][]{Anu06} was higher than the 1.7\,GHz flux density of
$36.7\pm0.3$\,mJy that we measured from the eastern part of the ring
on day 20.8.  Either
the eastern side of the ring contains a complicated mix of emission
and absorption, or the spectrum reflects a very unusual energy
distribution of relativistic particles.

  Assuming the emission is synchrotron -- a conclusion shared by
\citet{OBr06c} --  standard equipartition
arguments \citep[e.g.,][]{Bur59} suggest minimum energies in
relativistic particles and fields of
$\sim10^{41}\left(1+k\right)^{4/7}\rm\,ergs$, where $k$ is the
ratio of the energy in protons to that in electrons; $k\sim100$ is
observed locally, while $k\sim40$ is predicted for strong shocks
\citep[e.g.,][]{Bec05}.  The equipartition magnetic field at the
location of the peak flux density in the ring is of order
$0.03\left(1+k\right)^{2/7}\rm\,G$, comparable to previous results based on
much lower quality data \citep{Bod85,Tay89}.  The corresponding magnetic
pressure is 
$2.7\times10^{-5}$\,dyn\,cm$^{-2}$.  As expected, these values are
well below the total energy and pressure in the shock front.

\subsection{The Eastern Component}

\subsubsection{Emission Mechanism and Speed}

The steep spectral index ($\alpha \sim -0.67$) and high surface
brightness ($T_B\sim10^7\rm\,K$ at 1.7\,GHz) of the eastern component
indicate that the emission must be primarily synchrotron.
The minimum (equipartition) energy is then
$\sim0.15\times10^{41}\left(1+k\right)^{4/7}\rm\,ergs$, a factor 10
below that in the ring, while the corresponding equipartition magnetic
field is comparable, $\sim0.03\left(1+k\right)^{2/7}\rm\,G$.

If this component was ejected from the source at the core of the
expanding ring, it is moving quite rapidly.  Presumably it was ejected
during or after the initial outburst, and is thus at most 27\,days old
when first detected by us (2006 Mar 11).  At a distance of
$\sim52\rm\,mas$ from the center of the ring, its projected velocity
is $\gtrsim2\rm\,mas\,day^{-1}$, which corresponds to
$\gtrsim8500/\sin i\rm\,km\,s^{-1}$ (where $i$ is the inclination to
the line-of-sight, and taking a distance of 2.45\,kpc).  For an 
inclination $i\sim50\degr$, the velocity of the eastern component is 
$\sim11,000\rm\,km\,s^{-1}$.  For comparison, the escape speed from a
white dwarf is $\sim6530 (1-M/1.456\,M_\sun)^{-1/4}\rm\,km\,s^{-1}$
\citep{Han94}, or $\sim12,600\rm\,km\,s^{-1}$ for $M = 1.35\,M_\sun$.
Based on the contour plot of \citet{OBr06c}, the EVN detection on day
21.5 gives roughly the same speed.  The eastern component had thus not
decelerated significantly before day 27.

\subsubsection{First Appearance}

  \citet{OBr06c} present a preliminary view of VLBA, EVN, and MERLIN
observations of RS\,Oph in 2006, with the VLBI data covering days
13.8--28.7.  Most aspects of our images agree reasonably well with
their results.  However, they detect the eastern component quite
strongly (peak $\sim0.8\rm\,mJy\,beam^{-1}$) at 5\,GHz with the EVN on
day 21.5, while we do {\it not} see that component with the VLBA on
day 20.8.  This apparent discrepancy is likely due to the more
centrally concentrated uv-coverage of the EVN observations, which are
correspondingly more sensitive to faint, extended emission. The
eastern component was therefore already extended by day 20.8, as we
directly observe on day 26.8.  Further, our detection of the component
at 5\,GHz with the same instrument and observing time on day 26.8 as
our non-detection on day 20.8 shows that the eastern component was
still brightening during this time.

  Our 1.7\,GHz VLBA observations on day 20.8 do not show the eastern
component.
This result agrees with the non-detection of O'Brien et al. (2006) at
the same frequency on day 20.5.
Since the VLBA is much more sensitive to extended structures at this
lower frequency than at 5 GHz (the angular response scales as
$\lambda/B$, with $\lambda$ the wavelength, and $B$ the length of the
baseline under consideration), our non-detection at 1.7\,GHz places
strong constraints on the spectrum around day 21.
Had the spectrum on day 21 been as steep as the spectrum we
observed on day 27, we would have measured a 1.7\,GHz flux density of
$\gtrsim1.7$\,mJy, or $\gtrsim10\sigma$ in our VLBA observations.
Considering instead the $5\sigma$ upper limit of $0.6$\,mJy/beam at
1.7\,GHz, and comparing this limit to the EVN detection, we find that
on day 21, the radio spectrum was actually either flat or inverted
(rising towards higher frequencies).  The eastern-component spectrum
therefore evolved significantly between day 21 and day 27.


\subsubsection{Absorption}

  The spectral and flux density evolution, combined with the inferred
large extent of the eastern component, suggest that external
(free-free) absorption is important, as expected for a source embedded
in a dense, ionized red giant wind.  The primary challenge lies in
explaining the initial obscuration of the eastern component at a time
when the ring is clearly visible.  \citet{OBr06c} invoke a special
geometry to explain the visibility of the ring, as well as the late
turn-on of the eastern component and the lack of any corresponding
western feature during the first month.
In their picture, the obscuring medium is the circumstellar material
(assumed spherically symmetric) shed by the red giant companion.  They
assume the radio emission 
comes from a bright, slowly expanding equatorial ring and a bipolar
shock structure.
They take the approaching (eastern) side of the bipolar jet to be
inclined about $30\degr$ to the line-of-sight, moving outward, and
sweeping up absorbing material.
In this picture, the eastern component of the jet becomes visible when
the shock overtakes most of the absorbing circumstellar material.  The
receding jet remains behind more of the red giant wind, so that
(western) component is not expected to be seen until later. They
suggest that the more distant (eastern) part of the ring is visible
almost instantly because the jet has swept away the densest,
high-obscuration part of the red giant wind along that line-of-sight.

\begin{figure}
\plotone{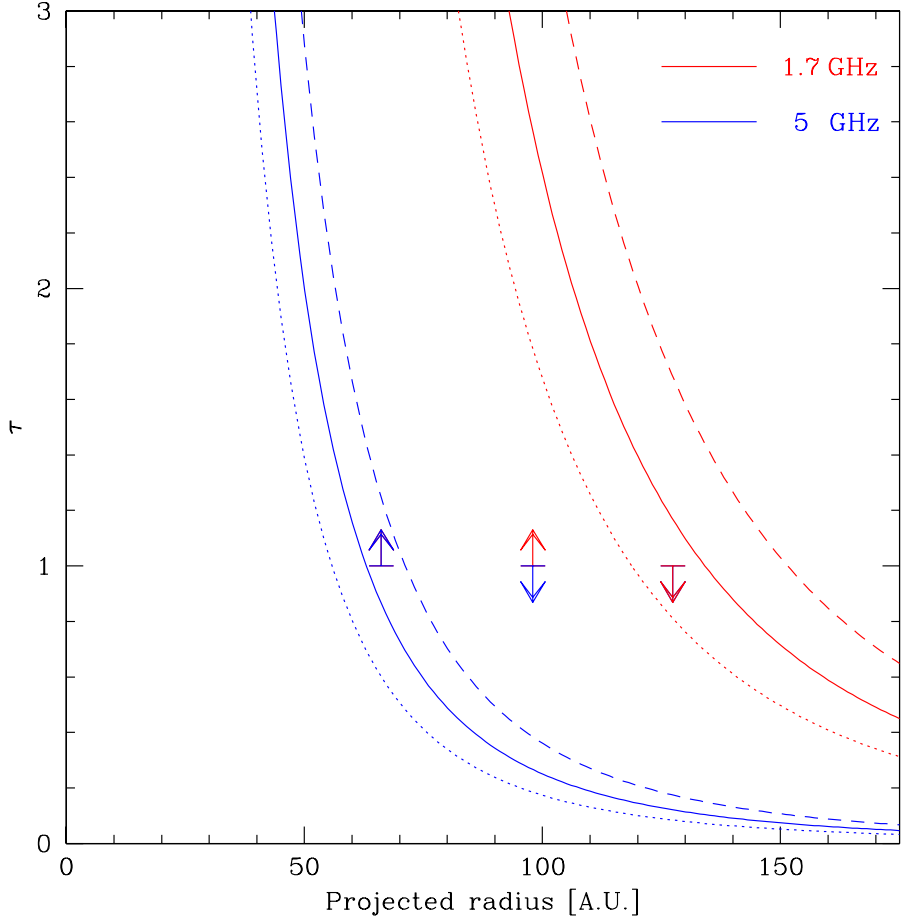}
\caption{The free-free opacity predicted for a $1/r^2$ wind from the
  red giant, at 5\,GHz (blue) and 1.7\,GHz (red), as a function of the
  (observed) projected distance from the red giant.
  The red and blue arrows
  show the rough constraints from the VLBI imaging of the eastern component ---
  the opacity must be $\lesssim1$ when that component is not
  detected, and $\gtrsim1$ when it is detected. 
  The horizontal placement
  of these limits assumes $D=2.45$\,kpc; a smaller distance would shift these
  to the left (implying a lower $\dot M/v_w$), while a larger distance would
  shift these to the right (implying a higher $\dot M/v_w$).
  The solid curves show a
  rough fit to these limits, corresponding to
  $\dot M/v_w= 1\times10^{-6}\,M_\sun\rm\,yr^{-1}/20\,km\,s^{-1}$ for
  an inclination of $i\sim50\degr$.
  The dotted lines show the opacities for a 20\%\ lower $\dot M/v_w$,
  while the dashed lines show the opacities for a 20\%\ higher value.
  These calculations take the electron temperature to be $10^4$\,K, and
  assume one proton per electron.
  \label{fig:abs}
  }
\end{figure}

  One can make this more quantitative by requiring that the free-free
opacity $\tau$ due to the red giant wind be $\gtrsim1$  at the times and
frequencies where the eastern component is not seen, and $\lesssim1$ at the
times and frequencies where it is.  The opacity predicted for a fully
ionized, spherically
symmetric, $1/r^2$ wind are shown as a function of projected distance from
the center in
Fig.\,\ref{fig:abs},  together with arrows showing the rough constraints
from the  1.7 and 5\,GHz VLBI observations.
This simple model works
remarkably well.  Assuming one proton per electron and an electron temperature
of $10^4$\,K, and taking the wind velocity
$v_w=20\rm\,km\,s^{-1}$, the inferred mass loss rate is
$\sim1\times10^{-6}\,M_\sun\rm\,yr^{-1}$ for an inclination $i\sim50\degr$
for the eastern component's  velocity vector.  This mass loss rate
is quite reasonable, and within a factor of two of that derived by
\citet{OBr92} (after correcting for the difference in the assumed distance)
based on shock models fit to the X-ray emission from the
1985 outburst.  

  Absorption can thus easily explain the late appearance of the eastern
component; but the central ring is more challenging.
In the \citet{OBr06c} model, the observed ring should
represent the projection of a highly-inclined, two-dimensional circle,
i.e., an ellipse.  Our data show that the ring in fact is quite
circular (Fig.\,\ref{fig:rings}), making explanations involving strong
deviations from spherical symmetry difficult to maintain.
The flat spectrum of the eastern part of the ring is also not a
natural result of the \citet{OBr06c} model.

  Another possible explanation for the abrupt appearance of the eastern
component is that it was intrinsically faint until around day 21.  
The eastern component 
could represent a
terminal shock of the kind associated with the ejecta of many active
galactic nuclei, microquasars, and protostars (in the form of Herbig
Haro objects), although this explanation has some difficulties since
for RS Oph, the
circum-binary density falls off rapidly with distance rather than abruptly
increasing.

The ejecta could
also have brightened upon moving into the
relative vacuum beyond the outer edge of the wind nebula.  As in the
current outburst, the expanding shock in 1985 swept up the CSM,
leaving behind a relative vacuum.  Over the past 21 years, the red
giant has been re-filling this vacuum with a wind moving out at
$\sim20\rm\,km\,s^{-1}$ \citep{Wal58,Duf64,Gor72}.  The collimated
ejecta would have
reached the edge of this re-filled region in $\sim14\rm\,days$
$\left(v_w /20\,{\rm km\,s^{-1}}\right)
\left(v_{jet}/11,000\,{\rm km\,s^{-1}}\right)^{-1}$ (where $v_{jet}$
is the velocity of the collimated outflow), around the
same time that it brightened in the radio.
If the red giant wind can carry outward and compress (and hence
strengthen) the magnetic field at large radii, relativistic particles
might produce copious synchrotron emission upon encountering the
compressed field.  Alternatively, if the relativistic plasma was
initially highly confined by the circumstellar gas (and thus
synchrotron self-absorbed), it might expand and become visible at
lower radio frequencies upon reaching the edge of the confining
medium.\footnote{Note that the azimuthal extent of the eastern
component requires a fairly wide-angle `jet'.  If the jet were
intrinsically narrow until reaching the edge of the red giant wind,
the azimuthal velocity would have to be $\gtrsim17,000\rm\,km\,s^{-1}$
to match the observed azimuthal extent on day 21.}  One might then
expect the eastern component to become relatively stationary at late
times rather than continuing to move out rapidly.

\section{Conclusions}

  Dual-frequency VLBA observations made 20.8 and 26.8 days after the
February 2006 outburst of the recurrent nova RS Oph have shown that
the bulk of the radio emission at this stage comes from an expanding
synchrotron shell.  The shell's increasing angular size matches the
shock velocities inferred from X-ray observations and measured through
infrared spectroscopy,
confirming the basic picture of a decelerating shock between the
material flung off in the explosion from the surface of the white
dwarf and the circumstellar material produced by the wind of its red
giant companion.  This gives a direct distance estimate of
$\sim2.45\pm0.05\pm0.37\rm\,kpc$, the most accurate yet determined for this source.

This evolving shock is far from simple.  The shell evolution
is {\it not} self-similar, with the shell thickness increasing from
$\sim15\%$ to $\sim26\%$ between the two epochs.  In addition, the
brightness distribution around the shell is highly asymmetric, with
the eastern side many times brighter than the western side.  The
eastern side of the ring also has a much flatter radio spectrum.  To
our knowledge, such a spectrum has been seen only once before for a
high-brightness resolved source, in the radio images of the expanding shell
associated with the
peculiar X-ray outburst of the emission-line star CI\,Cam
\citep{Mio04}.
Finally, the flux density of the western part of the shell is {\it
increasing} with time, in contrast to the optically-thin behavior of
the ejecta-driven shocks in physically similar systems such as
supernovae \citep[e.g.,][]{Wei02}.

Our VLBA images also reveal an additional eastern component that is
well separated from the dominant ring.  `Jets' (generally just
elongations or separate components in barely-resolved sources) have
been seen in previous outbursts of this 
and other
symbiotic stars.  The VLBI observations of RS
Oph constitute the first unambiguous evidence for a synchrotron jet
emanating from a white dwarf system.  The inferred speed is
close to the escape speed from a high-mass white dwarf, in accord with the
generally good agreement between directed outflow and escape speeds
for physical systems ranging from low-mass protostars to black hole
binaries and active galactic nuclei\footnote{In supernovae and
gamma-ray bursts, the relevant escape speeds are those of the compact
objects which are formed, not those of the progenitor stars.}.
Similarly, synchrotron emission is characteristic of fast jets, where
emission far from the powering source is generally assumed to derive
from an interaction between the flow and the surrounding medium.  The
association of a highly directed and a quasi-spherical outflow is more
unusual, but certainly not unprecedented; for instance, the jet model
for gamma-ray bursts, and their association with core-collapse
supernovae \citep[e.g.,][]{Mac99}, has brought the physical connection
between quasi-spherical and highly directed outbursts into much
greater prominence.  

The eastern component that was so prominent on day 27 appeared to
``turn on'' around day 21.
Although this turn-on could be due to decreasing absorption by the
circumstellar medium as the eastern component moved outwards, there
are several difficulties with this picture.  Intrinsic brightening,
possibly associated with the component reaching the edge of the region
re-filled by the red giant wind since the last nova outburst in 1985,
may also play a role in the turn-on.

The preferred east-west axis for the observed asymmetries is
consistent with VLBI observations on day 77 of the last (1985)
outburst, which showed a strong east/west extension \citep{Tay89}.
The persistent east-west asymmetry suggests that this preferred axis
reflects an intrinsic asymmetry in the binary system, rather than (for
instance) some statistical irregularity in a particular explosive
event.  The binary orbit provides one possible source of the
asymmetry. 
Although a highly anisotropic density structure around the binary
could collimate the outflow, with the `jet' naturally focused along
the axis of least resistance \citep[cf.][]{Llo93}, one would then
expect the observed ring to correspond to the high-density `waist' of
this structure.  The ring is in fact circular in projection and
brightest along the trajectory of the eastern component.
Whatever the collimation mechanism in RS Oph, we may for the first
time be seeing the formation of a white-dwarf jet, as well as the most
compelling demonstration to date of both its high initial speed and
the physical emission mechanism (synchrotron emission).


\acknowledgments

We are grateful to the VLBA scheduling committee and the array
  operations staff for the efforts which made these observations
  possible.  As with most research on bright variable stars, this work has
  greatly benefited from the regular monitoring (and quick reporting)
  of both amateur and professional variable star observers, generally 
  publicized through the American
  Association of Variable Star Observers (AAVSO) and the Variable Star
  Network (VSNet).  We are further grateful to the referee, Mike Bode, for
  a number of useful comments, as well as for organizing an exceptionally
  productive workshop on this source in the summer of 2007.
  J.L.S.\ is supported by an NSF Astronomy and Astrophysics
  Postdoctoral Fellowship under award AST~03-02055.  This research has
  made use of NASA's Astrophysics Data System; the SIMBAD database,
  operated at CDS, Strasbourg, France; and the USNOFS Image and
  Catalogue Archive, operated by the United States Naval Observatory,
  Flagstaff Station (\url{http://www.nofs.navy.mil/data/fchpix/}).




{\it Facilities:} \facility{VLBA ()}.




\clearpage

\begin{deluxetable}{ccccc}
\tablecaption{\label{tab:spind}}
\tablewidth{0pt}
\tablehead{
\colhead{} & \colhead{}
  & \colhead{$S_{1.7}$\tablenotemark{a}}
    & \colhead{$S_{corr, 5.0}$\tablenotemark{b}}
  & \colhead{} \\
\colhead{Epoch} & \colhead{Feature}
  & \colhead{$\rm(mJy~beam^{-1})$} & \colhead{$\rm(mJy~beam^{-1})$}
  & \colhead{$\alpha$\tablenotemark{c}} 
}
\startdata
2006 Mar 05 & Entire ring & $43.1\pm0.3$ & $32.2\pm0.6$  & $-0.27\pm0.02$ \\
 & Ring-E\tablenotemark{d} & $36.7\pm0.3$ & $29.5\pm0.6$  & $-0.20\pm0.02$\\
 & Ring-W\tablenotemark{d} & $\phn6.4\pm0.3$ & $\phn2.8\pm0.6$
   & $-0.76\pm0.20$\\
2006 Mar 11 & Entire ring & $41.3\pm0.2$ & $28.8\pm0.2$  & $-0.33\pm0.01$ \\
 & Ring-E\tablenotemark{d} & $33.3\pm0.2$ & $24.8\pm0.2$  & $-0.27\pm0.01$\\
 & Ring-W\tablenotemark{d} & $\phn8.0\pm0.2$ & $\phn3.8\pm0.2$
   & $-0.67\pm0.01$\\
 & E.C.\tablenotemark{d} & $\phn4.2\pm0.2$ & $\phn1.9\pm0.2$ & $-0.71\pm0.10$\\
\enddata

\tablenotetext{a}{Flux density at 1.7\,GHz.}
\tablenotetext{b}{Flux density at 5.0\,GHz, corrected for flux density
  missing as found by simulating 5.0\,GHz observations of the 1.7\,GHz model
  (see text). The observed flux densities have been multiplied by 1.07 on
  2006 March 05, 1.10 for the ring on 2006 March 11, and 1.21 for the eastern
  component on 2006 March 11.}
\tablenotetext{c}{Spectral index $\alpha$, with the flux density going
  as the frequency to this power ($S_\nu\propto\nu^\alpha$).}
\tablenotetext{d}{Ring-E refers to the eastern (bright) part, and Ring-W to
  the western (dim) part, of the ring. E.C.\ refers to the eastern
  component.}


\end{deluxetable}

\clearpage

\begin{deluxetable}{cccccc}
\tablecaption{\label{tab:radfits}}
\tablewidth{0pt}
\tablehead{
\multicolumn{2}{c}{Time covered by fit\tablenotemark{a}} & \hbox{\quad}
  & \multicolumn{3}{c}{Fit parameters\tablenotemark{b}}\\
\cline{1-2} \cline{4-6}
\colhead{$t_1$} & \colhead{$t_2$} &
  & \colhead{$R_{21}$\tablenotemark{c}} & \colhead{} & \colhead{} \\
\colhead{(days)} & \colhead{(days)} &
  & \colhead{(A.U.)} & \colhead{$\alpha$\tablenotemark{d}}
  & \colhead{$\chi^2/N_{d.o.f.}$\tablenotemark{e}}
}
\startdata
$-0.23\pm0.15$ & $\phn1.47$ && $\phn3.44\pm0.34\phn$ & $\equiv 0.0$ & \nodata\\
$\phn1.47$     & $14.0\phn$ && $15.49\pm0.17\phn$ & $-0.350\pm0.018$ &
  2.94/3 \\
$14.0\phn$     & $20.8\phn$ && $\phn5.70^{+0.039}_{-0.038}$ &
  $-0.524^{+0.022}_{-0.019}$ & 2.58/3\\
$14.0\phn$     & $26.8\phn$ && $\phn9.96^{+0.063}_{-0.056}$ &
  $-0.524^{+0.022}_{-0.019}$ & 2.58/3\\[0.2in]

$t_{expl}$\tablenotemark{f} & $20.8\phn$ && $24.63\pm0.38\phn$ &
  \nodata & \nodata \\
$t_{expl}$\tablenotemark{f} & $26.8\phn$ && $28.89^{+0.38\phn}_{-0.39\phn}$ &
  \nodata & \nodata \\
\enddata

\tablenotetext{a}{Times $t_1$ to $t_2$ over which the fit holds,
  measured in days with
  respect to the time of optical maximum $t_0\equiv$\,Feb.\,12.83.}
\tablenotetext{b}{Results of least-squares power-law fits to the X-ray
  data, parameterized in terms of the distance traversed by the shock
  $R_{21}$, and the power-law index $\alpha$. {\it Note that the distances
  have not yet been scaled to match the broad infrared spectral component}
  -- see \S\,\ref{sec:distance}.}
\tablenotetext{c}{The distance the shock has traveled between $t_1$ and
  $t_2$, in astronomical units (A.U.).  {\it Note that these distances have
  not yet been scaled to match the broad infrared spectral component} -- see
  \S\,\ref{sec:distance}.
  Error bars are $1\sigma$ based on
  standard propagation of errors, and are primarily statistical. The
  exception is the uncertainty in the explosion date ($t_{expl}=-0.23\pm0.15$),
  which is a best guess as to the likely range of dates.}
\tablenotetext{d}{The power-law index $\alpha$ of a fit to the velocities:
  $v\propto t^\alpha$. Error bars are $1\sigma$.}
\tablenotetext{e}{$\chi^2$ and the number of degrees-of-freedom for the fit.
  Note that $\chi^2$ has {\it not} been normalized --- $\chi^2\sim
  N_{d.o.f.}$ indicates a good fit.}
\tablenotetext{f}{These rows give the total distance traveled between
  the (uncertain) date of the explosion $t_{expl}$, and the dates of our
  VLBA observations (20.8 and 26.8 days).  The error bars are $1\sigma$
  based on a simple propagation of the error bars for each segment of the
  fit, assuming the error bars on the individual segments are independent.
  {\it Note that these distances have not yet been scaled to match the broad
  infrared spectral component} -- see \S\,\ref{sec:distance}.}


\tablecomments{The results given here have not yet been 
  scaled to match the velocities measured for the broad infrared spectral
  component -- see \S\,\ref{sec:distance}.}

\end{deluxetable}


\clearpage

\begin{thebibliography}{}
  
\bibitem[Anupama \& Kantharia(2006)]{Anu06} Anupama, G.~C.\ 
  \& Kantharia, N.~G.\ 2006, \iaucirc, 8687 
 
\bibitem[Bartel et al.(2002)]{Bar02} Bartel, N.\ et al.\ 2002,
  \apj, 581, 404

\bibitem[Bertout \& Magnan(1987)]{Ber87} Bertout, C.\ \& Magnan, C.\ 1987,
  \aap, 183, 319

\bibitem[Bode \&\ Kahn(1985)]{Bod85} Bode, M.~F.\ \&\ Kahn, F.~D.\ 1985,
  \mnras, 217, 205

\bibitem[Beck \&\ Krause(2005)]{Bec05} Beck, R.\ \& Krause, M.\ 2005,
  Astron.\ Nachr., 326, 414

\bibitem[Bode(1987)]{Bod87} Bode, M.~F.\ 1987, ed.,
  RS Ophiuchi (1985) and the Recurrent Nova Phenomenon, Utrecht: VNU Science
  Press

\bibitem[Bode et al.(2006)]{Bod06} Bode, M.~F., et al.\ 2006, 
  \iaucirc, 8675 

\bibitem[Bode et al.(2006b)]{Bod06b} Bode, M.~F., et al.\ 2006, 
  \apj, 652, 629 

\bibitem[Brandi et al.(2007)]{Bra07} Brandi, E., Quiroga, C., Ferrer, O.~E.,
  Mikolajewska, J., \&\ Garc\'\i a, L.~G.\ 2007, poster at the Keele
  University workshop on RS Ophiuchi 2006 (12-14 June 2007).

\bibitem[Brocksopp et al.(2003)]{Bro03} Brocksopp, C., Bode, 
M.~F., \& Eyres, S.~P.~S.\ 2003, \mnras, 344, 1264

\bibitem[Brocksopp et al.(2004)]{Bro04} Brocksopp, C., 
Sokoloski, J.~L., Kaiser, C., Richards, A.~M., Muxlow, T.~W.~B., \& 
Seymour, N.\ 2004, \mnras, 347, 430

\bibitem[Buil(2006)]{Bui06} Buil, C.\ 2006, CBET, 403

\bibitem[Burbidge(1959)]{Bur59} Burbidge, G.R.\ 1959, \apj, 129, 849

\bibitem[Chevalier(1982)]{Che82} Chevalier, R.A.\ 1982, \apj, 258, 790

\bibitem[Crocker et al.(2001)]{Cro01} Crocker, M.~M., Davis, 
R.~J., Eyres, S.~P.~S., Bode, M.~F., Taylor, A.~R., Skopal, A., \& Kenny, 
H.~T.\ 2001, \mnras, 326, 781
 
\bibitem[Das et al.(2006a)]{Das06a} Das, R.~K., Ashok, N.~M., 
\& Banerjee, D.~P.~K.\ 2006a, \iaucirc, 8673 
 
\bibitem[Das et al.(2006b)]{Das06b} Das, R., Banerjee, D.~P.~K., \&\ 
  Ashok, N.~M.\ 2006b, \apjl, 653, L141 
 
\bibitem[Davis(1987)]{dav87}
  Davis, R.~J. 1987, in 
  RS Ophiuchi (1985) and the Recurrent Nova Phenomenon (ed. M.F.\ Bode),
  Utrecht: VNU Science Press, 187

\bibitem[Dobrzycka \&\ Kenyon(1994)]{Dob94} Dobrzycka, D.\ \&\ Kenyon, S.J.\
  1994, \aj, 108, 2259

\bibitem[Dufay et al.(1964)]{Duf64} Dufay, J., Bloch, M., Bertaud, C., \&\
  Dufay, M.\ 1964, Annales d'Astrophysique, 27, 555

\bibitem[Evans et al.(2006)]{Eva06} Evans, A.\ et al.\ 2006, 
\iaucirc, 8682 

\bibitem[Evans et al.(2007)]{Eva07} Evans, A.\ et al.\ 2007, \mnras, 374, L1
 
\bibitem[Eyres et al.(2006)]{Eyr06} Eyres, S.~P.~S., O'Brien, 
T.~J., Muxlow, T.~W.~B., Bode, M.~F., \& Evans, A.\ 2006, \iaucirc, 8678 

\bibitem[Fey et al.(2004)]{Fey04} Fey, A.L.\ et al.\ 2004, \aj, 127, 3587

\bibitem[Fomalont et al.(2003)]{Fom03} Fomalont, E., Petrov, L., McMillan,
  D.S., Gordon, D., \&\ Ma, C.\ 2003, \aj, 129, 1163

\bibitem[Galloway \&\ Sokoloski(2004)]{Gal04} Galloway, D.~K.\ \&\
  Sokoloski, J.~L.\ 2004, \apjl, 613, L61

\bibitem[Gill \& O'Brien(1999)]{Gil99} Gill, C.~D.\ \& O'Brien, T.~J.\ 1999,
  \mnras, 307, 677

\bibitem[Gonzalez-Riestra et al.(2006)]{Gon06} 
Gonzalez-Riestra, R., Orio, M., \& Leibowitz, E.\ 2006, \iaucirc, 8695 

\bibitem[Gorbatskiii(1972)]{Gor72} Gorbatskii, V.G.\ 1972, 
  Soviet Astronomy, 16, 32

\bibitem[Greisen(2003)]{Gre03} Greisen, E.W.\ 2003, in
   Information Handling in Astronomy -- Historical Vistas, ed.\ A.\ Heck
   (Dordrecht: Kluwer Academic Publishers), 109

\bibitem[Hachisu \& Kato(2001)]{Hac01} Hachisu, I.\ \& Kato, 
  M.\ 2001, \apj, 558, 323
 
\bibitem[Hansen \& Kawaler(1994)]{Han94} Hansen, C.~J.\ \& Kawaler, S.~D.\
  1994, Stellar Interiors, New York: Spriunger-Verlag

\bibitem[Hjellming et al.(1986)]{Hje86} Hjellming, R.~M., van 
Gorkom, J.~H., Seaquist, E.~R., Taylor, A.~R., Padin, S., Davis, R.~J., \& 
Bode, M.~F.\ 1986, \apjl, 305, L71


\bibitem[Hog et al.(2000)]{Hog00} Hog, E.\ et al.\ 2000,
  \aap, 355, L27

\bibitem[Hollis et al.(1997)]{Hol97} Hollis, J.~M., Pedelty, J.~A., \&\
  Kafatos, M.\ 1997, \apj, 490, 302

\bibitem[Iijima(2006)]{Iij06} Iijima, T.\ 2006, \iaucirc, 
8675 

%
\bibitem[Iijima \&\ Essenoglu(2003)]{Iij03}
   Iijima, T.\ \&\ Essenoglu, H.~H. 2003, \aap, 404, 997

\bibitem[Kafatos et al.(1989)]{Kaf89} Kafatos, M., Hollis, 
J.~M., Yusef-Zadeh, F., Michalitsianos, A.~G., \& Elitzur, M.\ 1989, \apj, 
346, 991

\bibitem[Kawabata et al.(2006)]{Kaw06}
  Kawabata, K.~S., Ohyama, Y., Ebizuka, N., Takata, T.,
  Yoshida, M., Isogai, M., Norimoto, Y., Okazaki, A., \&\ 
  Saitou, M.~S. 2006, \aj, 132, 433

\bibitem[Kellogg et al.(2001)]{Kel01} Kellogg, E., Pedelty, 
J.~A., \& Lyon, R.~G.\ 2001, \apjl, 563, L151
 
\bibitem[Kenyon(1986)]{ken86}
  Kenyon, S.~J. 1986, The Symbiotic Stars, Cambridge: Cambridge
  University Press

\bibitem[Lane et al.(2007)]{Lan07} Lane, B.F.\ et al.\ 2007, \apj, 
  658, 520
 
\bibitem[Lloyd et al.(1993)]{Llo93} Lloyd, H.~M., Bode, 
M.~F., O'Brien, T.~J., \& Kahn, F.~D.\ 1993, \mnras, 265, 457
 
\bibitem[Lynch et al.(2006)]{Lyn06} Lynch, D.~K.\ et al.\ 2006,
  \apj, 638, 987

\bibitem[MacFadyen \&\ Woosley(1999)]{Mac99} MacFadyen, A.I.\ \&\ Woosley,
  S.E.\ 1999, \apj, 524, 262

\bibitem[Mioduszewski \&\ Rupen(2004)]{Mio04} Mioduszewski, A.J.\ \&\ Rupen,
  M.P.\ 2004, \apj, 615, 432

\bibitem[Munari \&\ Zwitter(2002)]{Mun02} Munrai, U.\ \&\ Zwitter, T.\ 2002,
  \aap, 383, 188

\bibitem[Monnier et al.(2006)]{Mon06} Monnier, J.D.\ et al.\ 2006, \apjlett,
  647, L127

\bibitem[Narumi et al.(2006)]{Nar06} Narumi, H., Hirosawa, 
K., Kanai, K., Renz, W., Pereira, A., Nakano, S., Nakamura, Y., \& 
Pojmanski, G.\ 2006, \iaucirc, 8671 
 
\bibitem[Ness et al.(2006)]{Nes06} Ness, J.-U., et al.\ 2006, 
\iaucirc, 8683 
 
\bibitem[Nichols et al.(2007)]{Nic07}  Nichols, J.\ et al.\ 2007, \apj,
  660, 651

\bibitem[O'Brien, Bode, \&\ Kahn(1992)]{OBr92} O'Brien, T.~J., 
  Bode, M.~F., \& Kahn, F.~D.\ 1992, \mnras, 255, 683 
 
\bibitem[O'Brien et al.(2006a)]{OBr06a} O'Brien, T.~J., Muxlow, 
T.~W.~B., Garrington, S.~T., Davis, R.~J., Eyres, S.~P.~S., Bode, M.~F., 
Porcas, R.~W., \& Evans, A.\ 2006, \iaucirc, 8684 
 
\bibitem[O'Brien et al.(2006b)]{OBr06b} O'Brien, T.~J., Muxlow, 
T.~W.~B., Garrington, S.~T., Davis, R.~J., Porcas, R.~W., Bode, M.~F., 
Eyres, S.~P.~S., \& Evans, A.\ 2006, \iaucirc, 8688 
 
\bibitem[O'Brien et al.(2006c)]{OBr06c} O'Brien, T.~J., Bode, M.~F.,
  Porcas, R.~W., Muxlow, T.~W.~B., Eyres, S.~P.~S., Beswick, R.~J.,
  Garrington, S.~T., Davis, R.~J., \& Evans, A.\ 2006, \nat, 443, 279 
 
\bibitem[Osborne et al.(2006a)]{Osb06a} Osborne, J., et al.\ 
2006a, The Astronomer's Telegram, 764
 
\bibitem[Osborne et al.(2006b)]{Osb06b} Osborne, J., et al.\ 
2006b, The Astronomer's Telegram, 770
 
\bibitem[Padin, Davis, \&\ Bode(1985)]{pad85}
   Padin, S., Davis, R.~J., \&\ Bode, M.~F. 1985,
   \nat, 315, 306

\bibitem[Pojmanski(2006)]{Poj06} Pojmanski, G.\ 2006, \iaucirc, 8671

\bibitem[Pooley et al.(2006)]{Poo06} Pooley, G.~G., 
Sokoloski, J.~L., Rupen, M.~P., \& Mioduszewski, A.~J.\ 2006, The 
Astronomer's Telegram, 750

\bibitem[Porcas, Davis, \&\ Graham(1987)]{por87}
  Porcas, R.~W., Davis, R.~J., \&\ Graham, D.~A. 1987, in
  RS Ophiuchi (1985) and the Recurrent Nova Phenomenon (ed. M.F.\ Bode),
  Utrecht: VNU Science Press, 187

\bibitem[Quiroga et al.(2003)]{Qui03} Quiroga, C., Brandi, E., Ferrer, O.~E.,
  Garc\'\i a, L.~G., \&\ Mikolajewska, J.\ 2003, BAAA, 46, 35

\bibitem[Roeser \&\ Bastian(1988)]{Roe88} Roeser, S.\ \&\ Bastian, U.\ 1988,
  \aaps, 74, 449
 
\bibitem[Seaquist(1989)]{Sea89}
  Seaquist, E.~R.\ 1989, in Classical Novae (eds.\ M.~F.\ Bode \&\ A.~Evans),
  Chichester: Wiley, 143
%
\bibitem[Sokoloski et al.(2004)]{Sok04}
  Sokoloski, J.~L., Kenyon, S.~J., Brocksopp, C., Kaiser, C.~R., \&\
  Kellogg, E.M. 2004, in Revista Mexicana de Astronomia y Astrofisica
  Conference Series v.\ 20 (eds. G.\ Tovmassian \&\ E.\ Sion),
  35

\bibitem[Sokoloski et al.(2006)]{Sok06a} Sokoloski, J.~L., 
Luna, G.~J.~M., \& Mukai, K.\ 2006, The Astronomer's Telegram, 747
 
\bibitem[Sokoloski et al.(2006)]{Sok06b} Sokoloski, J.~L., 
Luna, G.~J.~M., \& Mukai, K.\ 2006, The Astronomer's Telegram, 754
 
\bibitem[Sokoloski et al.(2006)]{Sok06c} Sokoloski, J.~L., 
Luna, G.~J.~M., \& Mukai, K.\ 2006, The Astronomer's Telegram, 741
 
\bibitem[Sokoloski et al.(2006)]{Sok06d} Sokoloski, J.~L., 
Mukai, K., \& Luna, G.~J.~M.\ 2006, The Astronomer's Telegram, 737

\bibitem[Sokoloski et al.(2006)]{Sok06e} Sokoloski, J.L., Luna, G.J.M.,
  Mukai, K., \&\ Kenyon, S.J.\ 2006, \nat, 442, 276
 
\bibitem[Sostero \& Guido(2006)]{Sos06a} Sostero, G., \& 
Guido, E.\ 2006, \iaucirc, 8673 
 
\bibitem[Sostero \& Guido(2006)]{Sos06b} Sostero, G., \& 
Guido, E.\ 2006, \iaucirc, 8681 

\bibitem[Spoelstra et al.(1987)]{Spo87} Spoelstra, T.~A.~T., 
Taylor, A.~R., Pooley, G.~G., Evans, A., \& Albinson, J.~S.\ 1987, \mnras, 
224, 791

\bibitem[Tatischeff \& Hernanz(2007)]{Tat07} Tatischeff, V.\ \& Hernanz, M.\
  2007, \apjl, 663, L101

\bibitem[Taylor et al.(1989)]{Tay89} Taylor, A.~R., Davis, 
R.~J., Porcas, R.~W., \& Bode, M.~F.\ 1989, \mnras, 237, 81 
 
\bibitem[Waagen et al.(2006)]{Waa06} Waagen, E., Labordena, 
C., Pearce, A., Granslo, B., Otten, C., Mavrofridis, G., \& Muyllaert, E.\ 
2006, \iaucirc, 8688 
 
\bibitem[Wallerstein(1958)]{Wal58} Wallerstein, G.\ 1958, \pasp, 70, 537
 
\bibitem[Weiler et al.(2002)]{Wei02} Weiler, K.W., Panigia, N., Montes,
  M.J., \&\ Sramek, R.A.\ 2002, \araa, 40, 387

\bibitem[West(2006)]{Wes06} West, J.~D.\ 2006, \iaucirc, 8683 

\bibitem[Zacharias et al.(2004)]{Zac04} Zacharias, N., Monet, D.,
  Levine, S., Urban, S., Gaume, R., \&\ Wycoff, G.\ 2004,
  \baas, 205, 4815

\end{thebibliography}
\end{document}